\documentstyle[preprint,pre,aps]{revtex}
\tightenlines
\begin{document}
\title
{Bifurcation and chaos in the double well Duffing-
van der Pol oscillator: Numerical and analytical studies}
\author{A. Venkatesan and M. Lakshmanan }
\address{
Centre for Nonlinear Dynamics,\\ Department of Physics,\\ Bharathidasan University,\\
Tiruchirappalli - 620 024, India}
\maketitle
\date{}
\begin{abstract}
The behaviour of a driven double well Duffing-van der Pol (DVP) oscillator
for a specific parametric choice ($\mid \alpha \mid  =\beta$) is studied. The
existence of different attractors in the system parameters ($f-\omega$)
domain is examined and a detailed account of various steady states for
fixed damping is presented. Transition from quasiperiodic to periodic
motion through chaotic oscillations is reported. The intervening chaotic
regime is further shown to possess islands of phase-locked states and
periodic windows (including period doubling regions), boundary crisis, all  the three classes  of intermittencies,
and transient chaos. We also observe the existence of local-global bifurcation
of intermittent catastrophe type and global bifurcation of blue-sky catastrophe type during transition
from quasiperiodic to periodic solutions. Using a perturbative  periodic solution,
an investigation of the various forms of instablities allows one to
predict Neimark instablity in the  $(f-\omega)$ plane and eventually results
in the approximate predictive criteria for the chaotic region.
\end{abstract}

\section{INTRODUCTION}
The Duffing-van der Pol oscillator (DVP)
\begin{equation}
\ddot x-\mu(1-x^2)\dot x+\alpha x+\beta x^3= f \cos \omega t,\mu>0,
\left({d\over dt}=.\right)
\end{equation}
is a ubiquitous nonlinear differential equation which makes its presence in physical,
engineering and even biological problems [1-7]. It is a generalization
of the classic van der Pol oscillator equation. It can be considered in
atleast three physically interesting situations, wherein the potential
$V(x)={ \alpha x^2 \over 2}+{\beta x^4 \over 2}$ is a  (i)   single
well $ ( \alpha  >0, \beta >0)$, (ii)  double well $(\alpha <0, \beta >0)$
or a (iii) double hump $(\alpha >0, \beta <0)$. Each one of the above
three cases has become a classic central model to describe inherently
nonlinear phenomena, exhibiting a rich and baffling variety of regular
and chaotic motions.

Chaotic motion in system (1) with a single well type restoring force was
investigated by Ueda and Akamatsu [8] as a model of negative
resistance oscillator and later on was studied by a number of other authors [9-12],
who noted symmetry breaking of attractors and the onset of chaotic dynamics.
Bountis etal.[11] have investigated the nonintegrability of a family of DVP
oscillators by studying analyticity properties of the solution in the
complex time plane and proved that infinitely sheeted structure (ISS) exists
in this system. Rajasekar, Parthasarthy and Lakshmanan [13] pointed out
that the DVP oscillator with double well potential exhibits Smale-horse shoe
chaos when transverse intersections of the homoclinic orbits occur. Further
Kao and Wang [14] had analog simulated the DVP oscillator with a double
hump potential and discussed the various mode locking, multiple hysteresis,
period doubling route to chaos, intermittent hopping and crises phenomena.

Recently Szemplinska-Stupnika and Rudowski [12] reported that a \underline {single
well} type DVP oscillator exhibits chaotic motion between two types of
regular motion, namely periodic and quasi periodic oscillations, in the
prinicipal resonance region for a specific values of the parameter $f(=1.0)$ and
a range of $\omega$ values (0.8 - 1.0). Also they have obtained a perturbative
solution for the periodic oscillation and carried out stability analysis
of such solution to predict Neimark bifurcation. However  no such analysis
exists for the important case of double well type DVP oscillator so far in
the literature, which is atypical in the sense that even in the absence of
forcing it shows the existence of multiple attractors [6,7].

Considering the DVP oscillator with a \underline {double well} type
restoring force in the form
\begin{equation}
\ddot x-\mu(1-x^2)\dot x-\mid \alpha \mid x+\beta x^3= f \cos \omega t,
\beta>0
\end{equation}
we notice that the three equilibrium points of the system (2) for $f=0$
correspond to $- \mid \alpha \mid x+ \beta x^3=0$, so that we have the stable fixed
points $x_{1,2}^{(s)}= \pm \surd {\mid \alpha \mid \over \beta}$ and the unstable fixed
point which is hyperbolic at $x_o^{(u)}=0$. Actually $x_{1,2}$ are elliptic
points for $\mid \alpha \mid= \beta$ and become stable foci for $\mid \alpha \mid>\beta$ while they are
unstable foci
for $\mid \alpha \mid<\beta$. As a result, the system (2) exhibits a large orbit (LO)
motion, which  always encircles all the three equilibrium points for the
case $\mid \alpha \mid=\beta$. As far as $\mid \alpha \mid >\beta$ is concerned, the system
exhibits both small orbit (SO), that is oscillation around any one of the
stable fixed points, and large orbit motion (LO), depending upon the values of the
other control parameters and also initial conditions .

In this paper we undertake an investigation of the dynamics of the double-well
DVP oscillator (2) and show that it is a rich dynamical system, possessing a
vast number of regular and chaotic steady states. In particular, considering
the special case $\mid \alpha \mid=\beta$ (the case $\mid \alpha \mid \not= \beta$ is even more richer
and the results will be presented separately [16]), we bring out the existence
of transition from quasiperiodic to periodic motion in the $(f-\omega)$
parameter space \underline {via chaotic motion}. The novel features we observe are that
in the chaotic sea there are many isles of periodic and phase-locked states,
which exhibit period doubling  phenomenon, intermittencies, crises etc., along with regions
of transient chaos, corresponding to local bifurcations. Besides, there are
also transitions from quasiperiodicity to period T orbits corresponding to
global bifurcation of blue-sky and local-global bifurcation of intermittent catastrophes. We also
present a perturbative approach to the study of bifurcations which occur
near the principal resonance.  The analysis allows us to derive the algebraic
equations for the instability boundaries.

The plan of the paper is as follows. In sec. II we present the numerical
results for different steady states, bifurcation routes and chaos for
system (2). In sec. III we develop a perturbative solution and obtain expressions
for the stability regions and compare them with numerical results. Then, we
compare the results with the dynamics of the double well Duffing oscillator in sec. IV.
Finally, sec. V summarizes our results.

\section{NUMERICAL RESULTS}
Equation (2) is
numerically integrated using the fourth order Runge-Kutta algorithm with
adaptive step size with parameter values fixed at $\mid \alpha \mid $=0.5, $\beta$
=0.5,  and $\mu$=0.1 inorder to study the large orbit behaviour mentioned in the
introduction.  The transitions are also characterized by tracing the time 
evolutions, phase portrait, Poincar\'e map, Fourier spectrum and Lyapunov
exponents. For identifing different steady
states, the dynamical transitions are traced out by two different scanning
procedures: (1) varying $\omega$ at a fixed $f$ (frequency scanning) and (2)
varying $f$ at a fixed $\omega$ (amplitude scanning). The resulting phase
diagram in the $(f-\omega)$ parameter is shown in Fig.1. The diagram
covers the transition thresholds in the region of principal and super
harmonic resonances, $0.4<\omega<1.0$ and the forcing strength lying in the
region $0.0<f<0.2$. The various features in the phase diagram are summarised
and the dynamical transitions of the attractors are elucidated in the
following.

\subsection{Phenomena of steady states}

One observes that equation (2) admits the free-running solution when the
external force is absent $(f=0.0)$.When it is present and for low $f$ values
and low $\omega$ values, the frequency of the system becomes incommensurate
with the external frequency. Consequently, the system exhibits multifrequency
quasiperiodic oscillations. When the value of the external frequency $\omega$
exceeds certain critical value for fixed $f$, a transition from quasiperiodic
to periodic oscillations occur on increasing $\omega$ (see Fig.1) essentially
due to supercritical Neimark bifurcation (see Sec. II B below and Sec III).
This phenomenon continues until a critical $f$ value $(f\sim 0.115)$.

On a further increase of the forcing parameter $f$, $f>0.115$, the system
exhibits chaotic motion between the two regular motions, that is
quasiperiodic and periodic oscillations, within a range of the driving
frequency $\omega$. For example, at $f=0.13$, chaotic motion occurs in
the region $\omega \in (0.546,0.553)$ (see Fig.2a).

As the forcing parameter $f$ increases further, within a very narrow
frequency region, chaos-periodic windows-chaos type transition is found to
occur between the two regular oscillations. For example at $f$=0.14, period
5T solution occurs in the frequency range $\omega \in (0.545,0.551)$ within
the chaotic range $\omega \in (0.525,0.553)$ (see Fig.2b). At $f$=0.17,
period doubling phenomenon occurs in the window region of the frequency
$\omega \in (0.503,0.529)$.  In the above range, we note that the period
2 orbit is born at $\omega=0.529$;it undergoes a period $2 \times 2^m$ doubling
bifurcations, finally leading to the onset of chaos as $\omega$
decreases (see Fig.2c).

Higher values of the forcing strength $f$ introduces the appearance of a
new transition. When this increase is coupled with increasing frequency,
the quasiperiodic motion suddenly changes into a phase locked attractor.
As an example at $f=0.19$, the transition from quasiperiodic oscillation to
phase locked states of period 3  orbit occur at $\omega=0.453$ and this locked state
persists in the frequency range $\omega \in (0.453,0.47)$, which is then
followed by chaos, reverse period doubling, chaos and periodic solution
(see Fig.2d).

The details of our  numerical study are summarised in the $(f- \omega)$
 phase diagrams, Fig.1 and Fig. 2. It depicts
the system parameter region where quasiperiodic, large periodic and chaotic
attractors exist.  The curves denoted QP(1),QP(2) and P(1) are the boundaries
of transition from quasiperiodic to the chaotic state, quasiperiodic to
periodic state and chaotic to periodic state respectively. Also one observes
in the entire transition regions where coexistence of multiple attractors occur.
Further, beyond the curve P(1), there are some regions which exhibit phase locked
periodic and  transient chaotic states. However, in this paper, those states are not discussed in  detail.

The various steady states as 1,2,3,...,15 denoted in Fig. 1 are then
illustrated in Fig.3.  Regular attractors are illustrated by their phase
portraits and quasiperiodic and chaotic attractors by their Poincar\'e
maps.

The first three points (1)-(3) in Fig.3 are examples of almost periodic
(quasiperiodic) orbits for low values of $f$. Then the points (4)-(15) are
essentially located at the principal and super harmonic resonance regions
for large values of the forcing parameter, $f>0.12$, at increasing driving
frequency $\omega$. We observe here the following: quasiperiodic orbit
(point (4)), period 3T orbit (point (5)), chaotic orbits (points (6),(7)),
period doubled orbits (points (8),(9)), chaotic orbits (points (10),(11)),
period 5T orbit (point (12)) and period T orbits (points (13)-(15)).

\subsection{Classification of Bifurcations}

The complicated dynamical behaviours of the DVP oscillator (2) with
$\mid \alpha \mid=\beta$ due to the presence of the double well restoring force
has been confirmed by the phase diagram as discussed above.  From the
bifurcation theory point of view, these correspond to several types of
bifurcations: secondary Hopf, intermittent
and blue-sky catastrophes besides standard period doubling bifurcations
which are discussed in the following section.

\subsubsection { QP(1) region - Local bifurcations: Secondary Hopf (Neimark) bifurcation}

In analogy with the Hopf bifurcation, a bifurcation is expected at a critical
value  as the limit cycle loses its stability, so that an attracting torus is
born. This is the secondary Hopf bifurcation or a Neimark bifurcation [17].
Further, the bifurcated solution can be either stable and supercritical or
unstable and subcritical. For the present DVP oscillator (2) with $\mid \alpha \mid=
\beta$, there is a very large transition region QP(1) corresponding to this
secondary Hopf bifurcation.  As an example, let us examine the transient
process near $f=0.1$.  Figures 4 show the Poincar\'e maps with
values of $\omega$ decreasing  and with the starting values of $x$ and $\dot x$
indicated as in brackets. In Fig.4a, for $\omega=0.59$, we see a node like
convergence to a point, while in Fig.4b for $\omega=0.58$ the convergence
has a spiralling character and the rate of convergence is noticebly slower.
In the last diagram of Fig.4c, at $\omega=0.57$, we see that the system
is moving outwards from near unstable fixed point towards the attracting
invariant closed curve and the approach is termed as supercritical. This
is the typical behaviour for the curve QP(1) in Fig.1 for $f<0.10$.

The other nature of Neimark bifurcation, namely subcritical behaviour has also been
observed in system (2). For example, at $f=0.15$, chaotic region exists
in the range $\omega \in (0.512,0.551)$ (see Fig.5). It follows that this narrow
strip of  chaotic motion is related to the transition from quasiperiodic
to periodic oscillation via chaotic motion. This transition does not occur
in a smooth, continous way as it is the case when the Neimark bifurcation
is a supercritical one but occurs through a chaotic region. This type
of transition has a close resemblence with the Duffing oscillator [15,17,18]
case where it was shown that a lower frequency band of chaotic region is
related to the saddle-node bifurcation, which causes a sudden change from/to
T-periodic orbit to/from chaotic attractor and transient motions separate
the two different steady states. Thus by analogy the occurrence
of chaotic motion in the region of driving frequency that separates
quasiperiodic and periodic oscillations  can be interpreted as a
subcritical Neimark bifurcation. This type of bifurcation has been reported
earlier in ref[12] for the single well DVP oscillator, but the unstable
motion corresponds to randomly transitional motion, whereas in the present
double-well case this corresponds to a fully chaotic attractor and periodic
windows.

\subsubsection {QP(2) region - Local-global bifurcation:Intermittent catastrophe}

Figs.6 show the typical Poincar\'e maps with forcing values $f=0.00172$
and 0.00173 at $\omega=0.83$ in the QP(2) region, where a transition from quasiperiodic
to periodic motion occurs. In order to lock the quasiperiodic motion to
the period T motion, the Poincar\'e map points form  a closed loop with
points progressing more rapidly near the top of the attractor and more
slowly where points are visibly dense. As the external force strength is
increased, a pair of saddle and node develops such that quasiperiodic
motion rapidly shrinks to the node, which is near the original attractor.
This type of process is prototypical of intermittent catastrophe [17,19].

\subsubsection {QP(2) region - Global bifurcations: Blue-sky catastrophe}

As the frequency value is increased to $\omega =0.998$, a different transition process
is captured for $f=0.05$. The typical Poincar\'e map is shown in Figs.7
with the values of $\omega$ at 0.998 and 1. and $f=0.05$. From Figs.7, we
find that a periodic attractor bifurcates to a quasiperiodic attractor which
is located inside  the other, and the two attractors are typically disjoint
and separated in phase space by a finite distance. Furthermore, it is not
generic for a quasiperiodic attractor to appear suddenly at the same control threshold
where periodic motion vanishes. The quasiperiodic attractor will have existed
previously, or it will not exist at all; in either case the bifurcation will
consist of periodic attractor simply losing stability - it vanishes
into the  blue. Such phenomenon is termed as blue-sky catastrophe [17,20] in the
literature and this event involves collisions with saddle type objects.

\subsubsection {V-shaped region - Transitions to chaotic attractors}

We now enumerate the various possible attractors present in the system in the
V-shaped region in Fig. 1.

\subparagraph*{(i) Transient chaos and boundary crises:}

Chaotic behaviour is observed between boundary of curves QP(1) and P(1) in Fig.1.
Near the boundary of each of the curves P(1), QP(1), the system behaves in a random way, with the
trajectory moving in phase space as if it were on a strange attractor.
However after a transient time, the motion settles into a regular attractor,
that is near P(1) it settles into a periodic motion ( for example see Fig.8) while near QP(1) into a
quasiperiodic oscillation. Such a phenomenon is termed as transient chaos, which is
a precursor to steady state chaos. Between these two transitional regions, periodic
windows, phase locked states, chaotic attractor and period doubling
phenomenon occur. In addition, the boundary crisis [21] of chaotic attractor
appears as the value of the  external frequency increases so that
the dynamics corresponds to the curve P(1), where the
boundary of the chaotic attractor touches the unstable periodic orbit. 

In addition, one observes the interesting fact that in the V-shaped region
there exists different parametric values for which intermittency of all the
three classic types occur.  This seems to be rare in such low dimensional systems. 

\subparagraph*{(ii) Type I Intermittency:}

The parameter regions separating the periodic windows inside the V- shaped region
correspond to various
complicated dynamics including chaos. The precise stability boundary of each
window  has been found to correspond to a saddle-node instability. As an example,
for $f=$0.17, if $\omega$ is increased across the saddle-node boundary,
type I intermittency occurs[22-25]. One such intermittency signature is shown in Fig.9.
The average laminar length $(<l>)$ of this type of intermittency is found to
comply with the law $<l>\sim \mu^{-\delta}$ with $\delta \sim 0.52 \pm 0.001$
where $\mu=\omega-\omega^c$ and $\omega^c $is the bifurcation threshold.

\subparagraph*{(iii) Type II Intermittency:}

In the earlier section, we showed that two possibilities exist in the QP(1) region
when the periodic motion encounters a Hopf-bifurcation.  Either quasiperiodic
motion results if the bifurcation is supercritical or a complicated evolution appears
if the
bifurcation is subcritical. In the later case the transition to chaos is found to have
intermittency signature in certain parametric regions.
Such an intermittency signature is shown in Fig. 10. A close look
into the signature reveals that there are distinct phases of the regular motion which  are
punctuated by other phases which are apparently chaotic. According to the
classical Pomeau-Manneville categorization of different types of intermittencies
based on local bifurcations, the present intermittency is of  type II since
the preceding bifurcation is a Hopf bifurcation [22-25].  The average length
$<l>$ of the laminar phase of this intermittency is found to comply with the
law $<l> \sim ({1 \over \mu})^{\delta}$  with $\delta$=0.9321
where $\mu=\omega-\omega^c$ and $\omega^c$ is
the bifurcation threshold.

\subparagraph*{(iv) Type III Intermittency:}
Next we discuss yet another type of route inwhich the periodic orbits in the
periodic windows are seen to undergo intermittent transition to chaos.
One such intermitteny
motion is shown in Fig. 11, which is a Poincar\'e time series plot for $f$=0.14 and
$\omega$=0.53802. It is seen that the motion just before the onset of
intermittency is of period 20 orbit which itself occurred due to the period doubling of period 10 orbit.
The laminar phase of Fig. 11  corresponds to period 40 orbit
along with chaotic bursts.
Therefore this is identified to have arisen out of a subcritical half subharmonic
instability,  that is subcritical period doubling. Thus according to PM  classification
this is type III intermittency [22]. The average length $<l>$ of the laminar
phase of this intermittency complies with the following  scaling law predicted
by Pomeau-Manneville: $ <l> \sim ({1 \over \mu})^{\delta}$ with  $\delta=0.9912$
where $\mu=\omega-\omega^c$ and $\omega^c$ is the bifurcation threshold.

\section{PERTURBATIVE ANALYSIS}

From the numerical studies reported in the earlier sections, we observed
that the T periodic orbit within a range of driving frequency $\omega$
for low values of $f$ is close to a harmonic function of time. Then by
obtaining a first order approximate period T $\left(= 2\pi\over{\omega}\right)$
solution and analysing the stability one can  estimate the system
parameters domain inwhich Neimark instability arises as was done in the case of
single well DVP oscillator for fixed $f$ and $\omega$ in ref [12]. Keeping
this aim in mind we look for a periodic solution of (2) 
using a perturbative method (with both $\mid \alpha \mid$ and $\beta$ fixed at 0.5).
Applying the method of multiple
scales [1,26] to equation (2), one can obtain the approximate solution
about the stable fixed point $x_s=\sqrt{\alpha \over \beta}$ in the form
\begin{equation}
x=-{3 \over 4} a^2 + a\cos(\omega t+\phi)+{ a^2 \over 4} \cos 2(\omega t+\phi)-
{ a^2 \over 3} \mu \sin 2(\omega t + \phi),
\end{equation}
where
\begin{equation}
a= {f \over \sqrt{\left(\Omega^2(a)-\omega^2\right)^2
+ \left({3 \over 2} \mu \omega a^2\right )^2}},
\end{equation}

\begin{equation}
\tan \phi= -{{3 \over 2} \mu \omega a^2 \over{\left(\Omega^2(a)-
\omega^2\right )}},
\end{equation}
and $\Omega^2(a)=1-{9\over8}  a^2 $ is
the natural frequency of the autonomous conservative system (2) at $\mu =0$
and $f=0$.
\subsection{Linear stability analysis}
\subsubsection{Soft-mode instability}

In order to examine the stability of the solution (3), we may look at a specific
form of instability which manifests itself by an 
exponential growth with time of the harmonic components in the solution (3). This can be
done by adding small disturbances to the amplitude and phase of the solution
(3) as 
\begin{eqnarray}
x&=&-{3 \over 4} (a+\delta a)^2 + (a+\delta a)\cos(\omega t+\phi+\delta \phi) +{ (a+\delta a)^2 \over 4} \cos 2(\omega t+\phi+\delta \phi)- \nonumber \\
& &{ (a+\delta a)^2 \over 3} \mu \sin 2(\omega t + \phi+\delta \phi).
\end{eqnarray}
Working out the linearized equation for $\delta a $ and $\delta \phi$, one
ultimatively arrives at the following expression which corresponds to the first
order  instability limit as
\begin{equation}
\omega ^4+\left({27 \over 4} \mu^2 a^4 -2 a^2-2 \right) \omega ^2
+\left(2 a^2+{53 \over 16} a^4 +1 \right)=0.
\end{equation}
The above analysis is valid only for a fluctuation having the same frequency
as the approximate solution x(t) considered.

\subsubsection{Hard-mode instability}

Now we examine another type of instability in which the perturbation may have different
harmonic components other than those in x(t).  Following the spirit of the work
of Szemplinska-Stupnicka and Rudowski [12], let us study the effect of a small
disturbance to $x(t)=\bar x(t)$, where $\bar x(t)$ is the solution (3), in the form
\begin{equation}
x(t)=\bar x(t)+\delta x(t).
\end{equation}
The linear variational equation for $\delta x(t)$ is then
\begin{equation}
\delta \ddot x+P_1(t)\delta\dot x+P_2(t)\delta x=0,
\end{equation}
where 
\begin{equation}
P_1(t)=-\mu \left[1-\left({a_0}^2+{a^2\over 2}+{a_2^2 \over 2}+{a_3^2 \over 2} \right)
+ 2a_0a \cos \theta 
+\left({a^2\over 2} \right) \cos 2\theta +... \right], \nonumber
\end{equation}
\begin{eqnarray}
P_2(t)&=& -\mid \alpha \mid+3\beta \left({a_0}^2+{a^2\over 2} + 2a_0a \cos \theta+{a^2\over 2} \cos 2\theta \right) \nonumber \\
     &  & +2\mu \left[\left({-a_0}^2a-{a^3 \over 4}\right)\omega \sin \theta +a_0a^2 \omega \sin 2 \theta - {a^3 \over 2} \omega \sin 3 \theta \right]+... 
\end{eqnarray}
and $a_0=-{3 \over 4}a^2$; $a_2={a^2 \over 4}$; $a_3={{a^3 \mu} \over 3}$;
$\theta=\omega t+\phi$, where dots correspond to higher harmonic components. Introducing now the
transformation
\begin{equation}
\delta x=u \exp \left[{-1\over 2} \int\limits_{0}^{t} P_1(t)dt\right],
\end{equation}
equation (9) can be converted into a Hill's equation
\begin{equation}
\ddot u+P(t) u=0,
\end{equation}
where
\begin{equation}
P(t)=P_2-{P_1^2\over 4}-{\dot P_1\over 2}. \nonumber
\end{equation}
Using the form of $P_1$ and $P_2$ given in (10) and (11), the transformation (12) can
be rewritten as 
\begin{equation}
\delta x=u \exp \left[-\Delta t +{\mu \over 2 } \int \limits_{0}^{t} \left(2a_0a \cos \theta
 +{a^2 \over 2} \cos 2 \theta \right) d\theta \right],
\end{equation}
\begin{equation}
\Delta ={1 \over 2} \mu  \left [ {a^2\over 2}+a_0^2+{a_2^2 \over 2}+{a_3^2 \over 2}
-1 \right].
\end{equation}
Applying the Floquet theorem, one can look for a particular solution
of (13) in the form
\begin{equation}
u=\exp (\epsilon_1 t).  \phi(t),
\end{equation}
where $\phi(t)$ is a periodic function of time and $\epsilon_1$ is either
real or imaginary. Thus the equation (15) becomes
\begin{equation}
\delta x(t)=\exp (\epsilon_1- \Delta) t. \bar \phi(t),
\end{equation}
where
\begin{equation}
\bar \phi(t)=\phi(t) \exp \left[{ \mu \over 2 } \int \limits_{0}^{t} \left(2a_0a \cos \theta
 +{a^2 \over 2} \cos 2 \theta \right) d\theta \right].
\end{equation}

The stability of solution (3) depends exclusively on the exponent coefficient
$(\epsilon_1-\Delta)$ in (18). Let now discuss the various possibilities
to have stable solution.

Case(i):  Considering  the case $\epsilon_1=\pm i\bar \epsilon_1$ so that
$\bar \epsilon_1$ is real and positive then
\begin{equation}
\delta x(t)=\exp (\pm i \bar \epsilon_1- \Delta) t. \bar \phi(t).
\end{equation}
It can be concluded that when
$\epsilon_1$ is imaginary, the solution (3) is stable if $\Delta>0$ or
a $>$ 0.91 and unstable if $\Delta <0$. This form of instability (termed as Neimark instability) leads
to a buildup of new harmonic components whose frequencies are incommensurate
with the frequency of the periodic solution (3).

Case(ii):  Considering  the case $\epsilon_1=\pm \bar \epsilon_1$ so that
$\bar \epsilon_1$ is real  then
\begin{equation}
\delta x(t)=\exp (\pm  \bar \epsilon_1- \Delta) t. \bar \phi(t).
\end{equation}
When  $\bar \epsilon_1$ is real,
the solution (3) is stable if $\Delta >0$ and $\Delta^2> \bar \epsilon_1^2$ and this
form of instability is approximately equal  to the classic first order instability as given
by Eq.(7).

Therefore the form of instability defined by the equation (20) and so
the condition $\Delta >0$ leads to the build up of new harmonic components
with frequencies  $\omega +  \bar \epsilon_1$ and  $\omega - \bar \epsilon_1$.
However, these frequencies are in general incommensurate with the frequency
$\omega$ of the periodic solution (3) whereas for $\Delta<0$ the solution
is unstable and so $\Delta=0$ is the boundary of the instability. Thus, this instability can be
interpreted as a Neimark instability, giving rise to a Neimark bifurcation.

Let us now look at the resonance curves, shown in Fig.12 and the two unstable
regions defined by the condition (7) (first order instability) and the
condition $\Delta <0 $ for Neimark instability. From the Fig.12, the Neimark
instability is expected to occur at the frequency value where the resonance
curve crosses the critical boundary value a $\sim$ 0.91.  
To determine the Neimark stability limit in the $(f-\omega)$ parameter
plane, we calculate the forcing parameter $f$ by using the resonance equation
(4). Fig.13  depicts the Neimark instability limit defined by the condition
$\Delta <0$ and the first order stability limit described by the condition (7).
The numerical study results presented already in Fig.1 are shown  
for comparison.
The theoretically   predicted Neimark instability values are  reasonably close to the
numerical results.

\section{Comparison in the dynamics with the double well Duffing Oscillator}

Finally it is of importance to compare the dynamics of the double-well
Duffing oscillator as given by Szemplinska-Stupnika and Rudowski [15],
\begin{equation}
\ddot x+\mu\dot x -\mid \alpha \mid x +\beta x^3 = f \cos \omega t,
\end{equation}
with that of double well DVP (2) for the same parametric values $\mid \alpha \mid=\beta=
0.5$ and $\mu=0.1$. The results are compared in Table I. In the lower frequency
boundary region (see Fig.1 of ref[15]), the Duffing oscillator exhibits symmetry breaking bifurcations while
in the DVP, Neimark bifurcation boundary(Fig.1 of the present paper) has been found. In the higher frequency
boundary region, particularly in the prinicipal resonance region, the Duffing system
exhibits period doubling bifurcations of small periodic orbits and cross well
chaos. However in the DVP system, this region is always found to have highly regular
orbits.

Further, the Duffing oscillator exhibits period doubling route to chaos of
large period T orbit in the lower frequency region, that is $\omega<0.4$, while
in the present case, the system always exhibits almost-periodic oscillations in
the region.  In the present case, in every transition boundaries such as QP(1),
P(1) and QP(2), there are some regions where coexistence of multiple attractors
are found to occur.  But in the Duffing oscillator case, the coexistence of
multiple attractors are observed in the transition region from large period T
orbit to cross well chaos. Blue-sky catasrophe, type II and III intermittencies  and
various phase locked states are found to occur in the present case. However, in the
Duffing oscillator case such phenomena have not been found (atleast to our
knowledge).  Naturally the dynamics exhibited by the DVP equation (2) is also
quite distinct compared to the forced van der Pol oscillator [27].

\section{Conclusions}

Numerical studies show that the double well Duffing-van der Pol oscillator (2)
with the parameter choice $\mid \alpha \mid = \beta$ exhibits a rich variety of
attractors of periodic, quasiperiodic and chaotic types. Four varieties of
transitions from quasiperiodic to periodic motions occur: (1) QP - periodic orbits
(2) QP - chaos - periodic,
 (3) QP - chaos - periodic windows - chaos - periodic and (4) QP - phase locked states - 
 chaos - periodic orbits. Besides these, local stable or
 supercritical, unstable or subcritical Neimark bifurcations and mode locking, intermittent
 catastrophe and blue-sky catastrophe bifurcations are also shown to exist.
 Transient chaos, period doubling phenomena, boundary crisis and intermittencies
 of all the three classic types are shown to occur and these were demonstrated with suitable
 examples in the ($f-\omega$) parameter space. In the literature, so far all the three
 intermittencies have been found to occur mostly in the higher dimensional or coupled systems [21-23]. However
 in the present case even in a single model of low-dimensional system, we are able to demonstrate all the three kinds of
 PM intermittencies.  The various forms of
 instabilities of approximate periodic solution allows one to predict the
 Neimark bifurcation in the ($f-\omega$) parameter domain. Although some
 discrepency between true and theoretical predictions occurs, the approximate
 analysis throws some light to distinguish between the regular and chaotic regions.

\section{Acknowledgements}
This work  forms part of a Department of Science and Technology,
Government of India research project. Also one of the authors (A.V) wishes to
acknowledge the Council of Scientific and Industrial Research, Government of India,
for providing a Senior Research Fellowship.

\begin{figure}
\caption[] {Regions of different steady states exhibited by the double well DVP
oscillator (2) at $\mu=0.1$.}
\label{Fig.1}
\end{figure}
\begin{figure}
\caption[]{(i) Bifurcation diagram for maximum amplitude x vs external forcing frequency
$\omega $, (ii) Maximal Lypanov
exponet $\lambda_{max}$ vs external forcing frequency $\omega$ of system (2).
(a) QP - chaos - periodic orbit transitions for $\omega \in $(0.4,0.65) at
$f$=0.13.
(b) QP - chaos - periodic windows - chaos transitions for $\omega \in $(0.4,0.65) at
$f$=0.14.
(c) QP - chaos - period doubling window - chaos transitions for $\omega \in $(0.4,0.65) at
$f$=0.17.
(d) QP - phase locked states - chaos - period doubling windows - chaos transitions
for $\omega \in $(0.4,0.65) at $f$=0.19.}
\label{Fig. 2}
\end{figure}

\begin{figure}
\caption[]{ Various types of steady states at $\mu=0.1$: (1)$f$=0.001, $\omega$
=0.78; (2) $f$=0.001, $\omega$=0.9; (3) $f$=0.12, $\omega$=0.45;
(4) $f$=0.15, $\omega$=0.45; (5) $f$=0.19, $\omega$=0.47; (6) $\omega$=0.515;
(7) $\omega$=0.52; (8) $\omega$=0.525; (9) $\omega$=0.53; (10) $f$=0.15, $\omega$=0.54;
(11) $f$=0.14, $\omega$=0.53; (12) $f$=0.135, $\omega$=0.54; (13) $f$=0.14, $\omega$=0.58;
(14) $f$=0.15, $\omega$=0.58; (15) $f$=0.18, $\omega$=0.58.}
\label{Fig. 3}
\end{figure}

\begin{figure}
\caption[]{Trajectories near the Neimark bifurcation for system (2): $f$=0.1:
(a) $\omega$=0.59(1,1); (b) $\omega$=0.59(1,1); (c) $\omega$=0.57(1,1).}
\label{Fig. 4}
\end{figure}

\begin{figure}
\caption[]{(i) Bifurcation diagram for maximum amplitude x vs  external forcing frequency $\omega $, 
(ii) Maximal Lypanov
exponet $\lambda_{max}$ vs external forcing frequency $\omega$ of system (2).
 QP-chaos-periodic orbit transitions for $\omega \in $(0.4,0.65) at
$f$=0.15.}
\label{Fig. 5}
\end{figure}

\begin{figure}
\caption[]{Poincar\'e map of the system (2) before and after mode locking for
(a) $f$=0.00172, (b) $f$=0.00173, (c) $f$=0.00174 at $\omega$=0.83.}
\label{Fig. 6}
\end{figure}

\begin{figure}
\caption[]{Poincar\'e map of blue-sky disappearence of Periodic attractor in system (2):
(a) $\omega$=0.998, (b) $\omega$=1. at $f$=0.05}.
\label{Fig. 7}
\end{figure}

\begin{figure}
\caption[]{Trajectories to show transient chaos for $f=0.125, \omega =0.535$.}
\label{Fig. 8}
\end{figure}

\begin{figure}
\caption[]{Signature of type I intermittency: Time series plot for $f$=0.17,
(a) $\omega$=0.52601; (b) $\omega$=0.526010001.}
\label{Fig. 9}
\end{figure}

\begin{figure}
\caption[]{Signature of type II intermittency: Time series plot for $f$=0.12,
(a) $\omega$=0.5532; (b) $\omega$=0.5537.}
\label{Fig. 10}
\end{figure}

\begin{figure}
\caption[] {Signature of type III intermittency: Poincar\'e time series plot for
$f$=0.14, $\omega$=0.53802.}
\label{Fig. 11}
\end{figure}

\begin{figure}
\caption[]{ Resonance curves and unstable regions of solution (3): I: branches
unstable in the sense of first order instability; II : branches unstable
in the sense of Neimark instability.}
\label{Fig. 12}
\end{figure}

\begin{figure}
\caption[]{Regions of different steady states : Numerical (solid line) and
theoretical (dashed line) stability limits.}
\label{Fig. 13}
\end{figure}
\begin{table}{\bf I. Comparison of the orbits of Duffing and DVP oscillators}\\
\begin{tabular}{llcll}
& Parameters &         & Duffing Oscillator & DVP Oscillator\\ 
           &         &                    &             \\ \hline
 & & $f$  small          & small orbit      & QP orbit\\
&$\omega \in$(0.4,0.6)& &&                       \\
&&$f$ large & large orbit and chaos &large orbit and chaos \\ \hline
 & & $f$  small          & small orbit &QP orbit\\
&$\omega \in$(0.6,1.0)& &&                       \\
&&$f$ large & small orbit and chaos &large orbit  \\ 
\end{tabular}
\end{table}

\end{document}